\documentclass[aps,pra,amssymb, amsfonts,amsmath,showpacs]{revtex4}
\begin{document}
\title{Completely Quantized Collapse and Consequences}
\author{Philip Pearle}
\email{ppearle@hamilton.edu}
\affiliation{Department of Physics, Hamilton College, Clinton, NY  13323}
\date{\today}
\begin{abstract}
{Promotion of quantum theory from a theory of measurement to a theory of reality requires an unambiguous specification of the ensemble of realizable states (and each state's probability of realization). Although not yet achieved within the  framework of standard quantum theory, it has been achieved within the framework of the Continuous Spontaneous Localization (CSL) wave function collapse model. In CSL, a classical random field $w({\bf x}, t)$ interacts with quantum particles.  The state vector corresponding to each $w({\bf x}, t)$ is a realizable state.  In this paper, I consider  a previously presented model,  which is predictively equivalent to CSL.   In this Completely Quantized Collapse (CQC) model, the classical random field is quantized.  It is represented by the operator $W({\bf x}, t)$ which satisfies $[W({\bf x}, t), W({\bf x}', t')]=0$. The ensemble of realizable states is described by a single state vector, the ``ensemble vector." Each superposed state which comprises the ensemble vector at time t  is the direct product of an eigenstate of  $W({\bf x}, t')$, for all ${\bf x}$ and for  $0\leq t'\leq t$, and the CSL state corresponding to that eigenvalue. These states never interfere (they satisfy a superselection rule at any time), they only branch, so the ensemble vector may be considered to be, as Schr\"odinger put it, a "catalog"  of the realizable states.  In this context, many different interpretations (e.g., many worlds, environmental decoherence, consistent histories, modal interpretation) may be satisfactorily applied.  Using this description, a long standing problem is resolved, where the energy comes from that particles gain due to the narrowing of their wave packets by the collapse mechanism. It is shown how to define energy of the random field and its energy of interaction with particles so that total energy is conserved for the ensemble of realizable states. As a byproduct,  since the random field energy spectrum is unbounded, its canonical conjugate, a self-adjoint time operator, can be discussed.   Finally, CSL is a  phenomenological description, whose connection to, or derivation from, more conventional physics has not yet appeared. We suggest that,  because CQC is fully quantized, it is a natural framework for replacement of the classical field $w({\bf x}, t)$ of CSL by a quantized physical entity.  Two illustrative examples are given.} 
\end{abstract}
\maketitle
\section{Introduction}\label{Section I}

	Quantum theory as a theory of measurement was promulgated by Bohr\cite{Bohr}, and this point of view has its adherents today\cite{Peres-Fuchs}.  But, the success of the theory suggests that it could be more, a theory of reality.  Einstein wrote Schr\"odinger\cite{Einsteinletter} that, if there were no more complete description of nature than that envisaged by Bohr, ``...then physics could only claim the interest of shopkeepers and engineers; the whole thing would be a wretched bungle."  Bell made much the same point, writing\cite{Bell} `` To restrict quantum mechanics to be exclusively about piddling laboratory experiments is to betray the great enterprise."
	
	The lack in standard quantum theory has been called the ``measurement problem," but I prefer to call it the ``reality problem."  I believe it is most precisely characterized as a failure to give a satisfactory response to the following operational requirement for a well-defined quantum theory of reality. You are given, at time $t=0$, a  state vector for an arbitrary physical system, together with its Hamiltonian, but no other information.  For $t>0$, specify (i.e., give a procedure for how to identify) the realizable states and their probabilities of realization.  
		
	Not only Bohr's version of quantum theory, but more recent sophisticated attempts, such as many worlds\cite{lev}, environmental decoherence\cite{zehzurek}, consistent histories\cite{griffithshartlegellmannomnes}, have not succeeded so far at providing such a specification\cite{mekent}.  They are successfully applied to many important and interesting examples, but these examples require additional information ({\it this} is the nature of the system,  {\it that} is the apparatus,  {\it there} is the environment...) which lie outside the theory.   
	
	The reality problem {\it has} been solved by enlarging quantum theory so that it describes state vector collapse as a physical process.  The Continuous Spontaneous Localization (CSL) model\cite{me, GPR, myreview}  provides a modified Schrodinger equation.   It describes how, in addition to their usual interactions, the particles of the world interact with a posited classical field $w({\bf x}, t)$ resulting in a nonunitary state vector evolution (which changes the norm of the state vector). What are the realizable states?  Each such state vector, corresponding to a different 
$w({\bf x}, t)$ (drawn from the class of white noise ``functions), is realizable.  What are the associated probabilities?  CSL provides a second equation, the prescription that the probability of occurrence of a state vector (the probability of occurrence of its associated $w({\bf x}, t)$) is proportional to its squared norm.  In this straightforward manner, the operational requirement cited above is satisfied.  Each of these CSL states accurately describes (except for occasional brief moments) the macroscopic world we see around us.  The predictions agree with all presently performed experiments (but there are testable differences from standard quantum theory).  The CSL model is reviewed in section II.  

	I  have previously presented a model\cite{Ways, mekent, myreview}  called the ``Index" model,  but  here  called  the ``Completely Quantized Collapse" (CQC) model.  It is reviewed in section III.  It utilizes a formalism invented to model continuous measurements\cite{contmeas}.   CQC is completely equivalent to CSL in its specification of the realizable states and their probabilities, in the following manner.
		
	In CQC, the classical field is replaced by a quantized field $W({\bf x}, t)$ which, like the classical field $w({\bf x}, t)$ of CSL, commutes with itself, $[W({\bf x}, t), W({\bf x}', t')]=0$.  There is a conjugate field, $\Pi({\bf x}, t)$,  which also commutes with itself, but which satisfies $[W({\bf x}, t), \Pi ({\bf x}', t')]=
i\delta(t-t')\delta({\bf x}-{\bf x}')$.  The basis of eigenstates of $W({\bf x}, t)$ for every ${\bf x}$, $t$, is denoted $|w\rangle$.   It satisfies $W({\bf x}, t)|w\rangle=w({\bf x}, t)|w\rangle$, i.e., there is a different eigenstate for every possible $w({\bf x}, t)$ function. There is a single state vector, called here the ``ensemble vector,"  which evolves unitarily (the Hamiltonian is given).   When the ensemble vector is expanded as a Schmidt decomposition in the ``preferred basis" $|w\rangle$, the particle state in the direct product with $|w\rangle$  is the CSL state associated with $w({\bf x}, t)$: schematically, 

\[ |\hbox{ensemble vector}, t\rangle=\sum_{\hbox{all}\thinspace\thinspace w}|w\rangle|\psi_{\hbox{csl}}, t\rangle_{w}.  
\]

\noindent  The usual quantum rule for probabilities (the squared norm of a state in the superposition) gives the CSL probability of realization of that state.  

	The measurement (reality) problem in standard quantum theory has often been phrased in terms of difficulties associated with the ill-defined collapse postulate, nowhere more succinctly than Bell's Boolean phrase``And is not Or," i.e., Schr\"odinger's equation gives a sum of terms but we observe one term or another.  CSL provides a resolution of the problem by individually, dynamically, providing each term we can observe.  However, CQC gives a sum of terms: how can that be considered as a resolution of the problem?   
	
	A colleague once observed to me that construction of CQC goes against everything I have worked for, the resolution of the reality problem by dynamical collapse.  However, what I have worked for is to make a well-defined quantum theory of reality\cite{Pearle'67}, i.e., one which satisfies the aforementioned operational requirement.  CSL satisfies this requirement, but so does CQC.  The terms in the superposition which make up the ensemble vector are precisely defined (by the orthogonal preferred basis $|w\rangle$), do not interfere (they obey a superselection rule), 
are macroscopically sensible (because the associated CSL states are), and occur with the correct amplitudes (which give the correct probabilities). In his famous ``cat paradox" paper\cite{Schr}, Schr\"odinger referred to the state vector as the ``prediction catalog" 
or the  ``expectation catalog" because, in Bohr's quantum theory, it lists what can be expected to occur if a measurement is performed.  The CQC state vector may, likewise, be called a ``reality catalog."

However, the justification here is greater than Schr\"odinger's.  His usage requires judgment {\it outside} the theory as to {\it when} the state vector has evolved sufficiently to be considered a catalog, and as to {\it which} are the appropriate states in the superposition (i.e., which basis) to consider as making up the catalog.   In CQC, these criteria lie {\it inside} the theory, since the ensemble-vector may {\it always} be considered a catalog, and the appropriate states are {\it labeled} by the preferred basis $|w\rangle$.  Roughly speaking, CSL provides a vertical catalog (Or), and CQC provides a horizontal catalog (And), but it is the same catalog.  
If one has a precise rule for identifying a superposition of states at time $t$ which never interfere thereafter,  it is a matter of indifference whether one removes them from the superposition and follows their separate evolutions that way, or leaves them in the superposition and follows their joint evolution that way.

As  previously pointed out\cite{mekent}, one may regard CQC as a model for a satisfactory resolution of other attempted interpretations of quantum theory\cite{GS}.  The superposed states making up the ensemble vector could be thought of as many worlds (although it would seem that there would then be no reason to insist upon such an interpretation).  If the $W$-field is regarded as the environment,  then CQC could be regarded  as  describing  environmental decoherence.  The evolving superposed states making up the ensemble vector could be taken to represent a single family of consistent histories\cite{kent}.   The Schmidt decomposition of the ensemble vector is precisely what the modal interpretation requires. What is currently missing from these approaches, and what CQC provides, is a subsystem with a well-defined ``universal basis" (here the $|w\rangle$ basis vectors) whose Schmidt decomposition partners make macroscopic sense.  One might hope that such a basis may be found arising in a natural way from a  physical entity, whence perhaps CSL/CQC and all these interpretations may merge (see the last two paragraphs of this introduction for further remarks on this). 

	Now I turn to the topics which will occupy the rest of this paper. Because collapse narrows wave functions,  the energy of particles in CSL increases with time.  This growth of energy of particles provides for experimental tests of the model\cite{exptlpapers} but, from a theoretical point of view, it is desirable to have a conserved energy.  Recently, I showed how to define energy associated with the $w$-field in CSL so that the first moment of the total energy (particles plus $w$-field) is conserved\cite{energy}.  However, energy conservation requires all moments of the energy to be conserved.  The formalism of CQC facilitates that demonstration (which can be converted to CSL language with less transparency).  Because the ensemble vector evolution in CQC is a Hamiltonian evolution, the Hamiltonian is a conserved energy associated with the ensemble. 

	The Hamiltonian does not commute with $W({\bf x}, t)$  since it is composed of the $W$ {\it and} $\Pi$-field operators. Thus one cannot associate a conserved energy with an individual realizable state labeled by eigenvalues of $W$ . Nonetheless, it is of interest to see how energy fares for the ensemble of  possible states, for example, how the ensemble energy of the $W$-field can go negative to compensate for the energy increase of the particles whose wavefunctions are narrowed by collapse.  The discussion of energy conservation takes place in section IV.

	Another example of the utility of the CQC model is considered in section V.   Because the  $W$-field energy has the whole real line for its spectrum, it is possible to construct its self-adjoint conjugate, a  time-operator $T$, built solely out of the $W$ and $\Pi$-field operators. One may consider that time should be intimately related to the realization of physical states, since without 	such a realization there are no events and without events there is no time.  This, and the fact that, like the Matterhorn, it is ``there" (i.e., very few viable physical theories possess an unbounded Hamiltonian) motivate looking at $T$.	
		
	$T$'s ensemble probability distribution is examined. $T$'s variance diminishes with time.   $T$'s mean, roughly speaking, is a ``center of time,"  i.e., the average time over all space, and over time from 0 to $t$, weighted by the square of the particle number density.  Thus, if the particle Hamiltonian vanishes so the only particle evolution is collapse, which does not change the ensemble spatial particle density, the mean of $T$ is $t/2$.  If, however, the particles have a high clumping rate (for example, due to gravitational interaction), the spatial average of the square of the particle number density is largest most recently, and the mean of $T$ is closer to $t$. 

	One of the motivations of this work is to provide a bridge from dynamical collapse models to other interpretations which seek a well-defined quantum theory based upon the standard quantum formalism alone (although they may be satisfied with just accounting for observation, not reality), rather than the altered  formalism of collapse models. Now, what I consider to be a pressing problem associated with CSL is that it is phenomenological. (Some have said ``ad hoc," but that is not appropriate terminology since ad hoc means ``for this case only," whereas CSL applies to all cases, all physical situations: in fact, the label ``ad hoc" more appropriately belongs to these other interpretations  to the extent that  they require additional information appropriate to each physical situation). It could be that CSL dynamics may arise from a larger structure which generates quantum theory itself, as has been suggested by Adler\cite{Adler}.  In this paper, we consider another possibility.
	
	CSL depends upon interaction of a posited field's interaction with particles, but no identification of that field with a physical entity has been made.  A purpose in discussing CQC  is that, since it is formulated completely quantum mechanically, that may make it easier to identify the universal $W$-field with a quantized physical entity which arises naturally in another context.  In that case, CQC would appear as based upon the standard quantum formalism alone.  
	
	However, it is not so easy to find a candidate for a universal fluctuating field with the desired properties. There is experimental evidence\cite{exptlpapers} that the coupling of  $W$ is to the mass-density of particles, which strongly suggests a gravitational connection, a proposal which has a long history\cite{gravity, pearlesquires}.  This sugggests looking for a mechanism whereby gravitational excitations could naturally act like the $W$-field.  

	By way of illustration, Section VI contains two models for $W$ with gravitational overtones, neither to be taken too seriously.   In one, $W$ is constructed out of many scalar quantum fields (of Planck mass), each interacting only for a brief interval  (the Planck time)  with the particle's mass-density.  The other utilizes aspects of a previous gravitational classical model\cite{pearlesquires} of $W$ and crudely pays homage to spin-network models\cite{smolinseth}.  It  assumes that space consists of Planck-volume cells, each containing a bound spin 1/2 entity with two possible energy states equal to $\pm$Planck mass.  A spin occasionally breaks free to briefly interact with a thermal bath, as well as gravitationally with the particle's mass-density, before it is bound once more.  In this case,  $W({\bf x},t)$ is identified with the free mean spin, in a small volume surrounding ${\bf x}$,  during the small time interval surrounding $t$,.  These examples suggest that what is required for $W$ is a quantity which 
has the freedom to vary independently over all space-time and for which the state vector describes its prior space-time values . This is unlike a conventional field, which can vary independently (in magnitude and time derivative) only on a single spacelike hypersurface, and whose present values are all that is described by the state vector.

\section{CSL}
\subsection{The Simplest Model}
	The simplest CSL model, which will supply most of the illustrative calculations in this paper, describes collapse toward an eigenstate of the operator $A$. The state vector evolution, governed by a function $w(t)$ of white noise class, is given by 
\begin{equation}\label{1} 
|\psi, t\rangle_{S}={\cal T} e^{-\int_{0}^{t}dt'\{iH_{A}+(4\lambda)^{-1}[w(t')-2\lambda A]^{2}\}}|\phi\rangle.
\end{equation}
\noindent In Eq.(1), ${\cal T}$ is the time-ordering operator, $H_{A}$ is the (usual) Hamiltonian of the system, $|\phi\rangle=|\psi,0\rangle$ is the initial state vector, the subscript $S$ denotes the Schr\"odinger-picture state vector, and the parameter $\lambda$ governs the collapse rate.  Eq. (1) is the solution of the modified Schr\"odinger equation:
\begin{equation}
d|\psi, t\rangle_{S}/dt=\{-iH_{A}-(4\lambda)^{-1}[w(t)-2\lambda A]^{2}\}|\psi, t\rangle_{S}.\nonumber
\end{equation}
\noindent However, only the solution form (1) shall be utilized in this paper.  

	CSL is completed by specifying the probability that a particular $w$, lying in the interval $(w(t), w(t)+dw),$ is realized in nature:	
\begin{equation}\label{2}
	P_{t}(w)Dw=_{S}\negthinspace\langle\psi,t|\psi,t\rangle_{S}Dw.
\end{equation}
\noindent  In Eq.(2), and for all integrations, one may discretize time, i.e., with $t_{j}\equiv jdt$, each $w(t_{j})$ is considered to be an independent variable ($-\infty<w(t_{j})<\infty$), and 
$Dw\equiv\prod_{j=1}^{t/dt}\{dw(t_{j})/[dt/2\pi\lambda]^{1/2}\}$, with the eventual limit $dt\rightarrow 0$.  

	In  the ``collapse interaction picture," which is what is mostly employed in this paper,  the state vector is 
\begin{subequations}
\begin{eqnarray}\label{3} 
|\psi, t\rangle&\equiv&\exp{iH_{A}t}|\psi, t\rangle_{S}\\
&=&{\cal T} e^{-(4\lambda)^{-1}\int_{0}^{t}dt'[w(t')-2\lambda A(t')]^{2}}|\phi\rangle, 
\end{eqnarray}
\end{subequations}	
\noindent where (3b) follows from Eq.(1), and $A(t')\equiv\exp(iH_{A}t)A\exp(-iH_{A}t)$. 
From Eq.(3a), the probability rule, Eq.(2), is the same, except that the subscript $S$ is dropped.   

	The density matrix describing the ensemble in the interaction picture follows from Eqs.(2), (3b):
\begin{subequations}\label{4} 
\begin{eqnarray}
\rho(t)&&\equiv\int P_{t}(w)Dw\frac{|\psi, t\rangle\langle\psi, t|}{\langle\psi, t|\psi, t\rangle}
=\int Dw|\psi, t\rangle\langle\psi, t| \\
&&={\cal T} e^{-(\lambda/2)\int_{0}^{t}dt'[A_{L}(t')-A_{R}(t')]^{2}}|\phi\rangle\langle\phi |
\end{eqnarray}
\end{subequations}	
\noindent where, in an expansion of the exponent in Eq.(4b), $A_{L}(t')$ $(A_{R}(t')$) appears to the left (right) of $|\phi\rangle\langle\phi |$, and is  time-ordered (reverse-time-ordered) by ${\cal T}$. 
\subsection{Proof of Collapse When $H_{A}=0$} 
	
	If $H_{A}=0$ (or if the collapse rate is so fast that, over the time interval $(0,t)$, we can well approximate $A(t)\approx A$), then the state vector dynamically collapses, as $t\rightarrow \infty$, to one of the eigenstates of A. Let $|\phi\rangle=\sum_{n}\alpha_{n}|a_{n}\rangle$, where 
$A|a_{n}\rangle=a_{n}|a_{n}\rangle$.  An intimation of the collapse behavior in this case may be seen from the density matrix (4b):
\begin{equation}
\rho(t)=\sum_{n,m} \alpha_{n}\alpha^{*}_{m}e^{-(\lambda t/2)(a_{n}-a_{m})^{2}}|a_{n}\rangle\langle a_{m}|,\nonumber
\end{equation}
\noindent since the off-diagonal elements vanish as $t\rightarrow \infty$, and the diagonal element magnitudes are the ones obtained from $|\phi\rangle$ using the ``collapse postulate" of standard quantum theory.  However, this ensemble density matrix behavior does not prove collapse in the individual case, since many different ensembles can have the same density matrix. 
 
	Here is a proof of collapse not previously given. Eqs.(3),(2) become respectively:
\begin{subequations}\label{5} 
\begin{eqnarray}	
|\psi, t\rangle&=&\sum_{n} \alpha_{n}e^{-(4\lambda)^{-1}\int_{0}^{t}dt'[w(t')-2\lambda a_{n}]^{2}}|a_{n}\rangle\\
P_{t}(w)&=&\sum_{n} |\alpha_{n}|^{2}e^{-(2\lambda)^{-1}\int_{0}^{t}dt'[w(t')-2\lambda a_{n}]^{2}}
\end{eqnarray}
\end{subequations}

	 It shall now be shown that the only $w(t)$'s of non-zero measure are those for which the asymptotic ($t\rightarrow\infty$) time-average of $w(t)$ is $2\lambda a_{n}$.  To see this, consider the joint probability of $w$ and $a\equiv (2\lambda t)^{-1}\int_{0}^{t}dt'w(t')$:
\begin{equation}\label{6} 
P_{t}(w,a)Dwda=2\lambda tda\delta\big[\int_{0}^{t}dt'w(t')-2\lambda ta\big]P_{t}(w)Dw,
\end{equation}
\noindent where $P_{t}(w)Dw$ is the probability of $w$ and $2\lambda tda\delta$ is the conditional probability, given $w$, that $w$'s time-average is $2\lambda a$.  To find the probability distribution  $P_{t}(a)$ of  $a$, one must integrate (6) over $Dw$.  To do so, make the replacement  
$\delta(z)=(2\pi)^{-1}\int d\omega \exp i\omega z$, and then the resulting gaussian integrals over $Dw$ 
may be performed.  Next, the (gaussian) integral over $\omega$ can be performed, resulting in:
\begin{equation}\label{7} 
P_{t}(a)da=\sum_{n} |\alpha_{n}|^{2}\frac{e^{-2\lambda t (a- a_{n})^{2}}}{\sqrt{\pi/2\lambda t}}da\quad_
{\overrightarrow{t\rightarrow\infty} }\quad\sum_{n} |\alpha_{n}|^{2}\delta (a- a_{n})da.
\end{equation}
\noindent  Thus,  the asymptotic ensemble probability is completely accounted for by $a$ taking on 
only the values $a_{n}$, and the probability of an $a_{n}$ is $|\alpha_{n}|^{2}$.  

	It remains to show that (5a) asymptotically describes collapse.  Let $w(t)$ satisfy the constraint  $C_{a}$ defined as $\lim_{T\rightarrow\infty}(2\lambda T)^{-1}\int_{0}^{T}dt'w(t')= a$.  Taking the asymptotic limit of (6) and then integrating over an infinitesimal range about $a$ followed by integrating over $w$, it follows from (6), (7) that 	
\[  |\alpha_{n}|^{2}\delta_{a,a_{n}}= 
 \int Dw\int_{a-\epsilon}^{a+\epsilon}da P_{\infty}(w,a)=\int_{C_{a}} Dw P_{\infty}(w).
\]
	
	Applying this result to (5b) yields
\[  \int_{C_{a}} Dw e^{-(2\lambda)^{-1}\int_{0}^{\infty}dt'[w(t')-2\lambda b]^{2}}=\delta_{a,b}.
\]
\noindent Since each term in the integral here is non-negative, if $a\neq b$, for {\it each} $w(t')$ obeying 
 the constraint $C_{a}$ (i.e., for {\it each} $w(t')$ which asymptotically equals $2\lambda a$ plus a 
 fluctuating term which averages out to 0), then $\ exp-(2\lambda)^{-1}\int_{0}^{\infty}dt'[w(t')-2\lambda 
 b]^{2}=0$.  
 
 	This result can be applied to Eq.(5a), where the gaussians are of the same form (except $4\lambda$ replaces  $2\lambda$).  Thus, for each such $w(t)$ which obeys the constraint $C_{a}$, with $a\neq a_{n}$, the state vector asymptotically vanishes:  that doesn't matter since, by Eq.(5b), the probability of its occurrence vanishes also.  When $w(t)$ obeys the constraint $C_{a_{m}}$, all terms  in (5a) but the $m$th asymptotically vanish, $|\psi,\infty \rangle\rightarrow\sim |a_{m}\rangle$: collapse occurs. 
 
\subsection{Full-Blown CSL}

	The CSL proposal for a serious collapse model, applicable to all nonrelativisitic systems, involves multiple $A$'s, each with its own $w$.  That is, collapse is toward the joint eigenstates of an operator $A({\bf x})$ for all ${\bf x}$, and $w(t)$ is replaced by the random field $w({\bf x},t)$.  The choice of $A$ is\cite{me,pearlesquires}:
\begin{equation}\label{8} 
A({\bf x})\equiv \sum_{i}\frac{m_{i}/m_{0}}{(\pi a^{2})^{3/4}}\int d{\bf z}e^{-(2a^{2})^{-1}({\bf x}-{\bf z})^{2}}\xi_{i}^{\dagger}({\bf z})\xi_{i} ({\bf z}).
\end{equation}
\noindent In Eq.(8), $\xi_{i} ({\bf z})$ is the annihilation operator at location ${\bf z}$ for a particle of mass $m_{i}$ ($m_{0}$ is the proton mass), and $a$ (and $\lambda$) are parameters from the seminal GRW model\cite{GRW}, with chosen values $a\approx 10^{-5}$cm and  $\lambda\approx 10^{-16}$sec$^{-1}$.  Thus,  since $\xi_{i}^{\dagger}({\bf z})\xi_{i} ({\bf z})$ is the particle number density of the $i$th particle type at ${\bf z}$, $A({\bf x})$ is essentially  proportional to the mass density inside a sphere of radius $a$ centered at {\bf x}.  The CSL dynamical equation which replaces Eq.(3b) is:
\begin{equation}\label{9} 
|\psi, t\rangle={\cal T} e^{-(4\lambda)^{-1}\int_{0}^{t}dt'd{\bf x}[w({\bf x},t')-2\lambda 
A({\bf x},t')]^{2}}|\phi\rangle
\end{equation}
\noindent (where $A({\bf x},t')\equiv\exp(iH_{A}t)A({\bf x})\exp(-iH_{A}t)$), i.e., time is replaced by space-time.  The probability rule is unchanged, $P_{t}(w)Dw=\langle \psi,t|\psi, t\rangle Dw$, but space-time is discretized for integration, so 
$Dw\equiv\prod_{j, k}\{dw({\bf x}_{k},t_{j})/[dtd{\bf x}/2\pi\lambda]^{1/2}\}$,

\section{CQC}

\subsection{Simplest Model}

	Start by introducing creation and annihilation operators $b^{\dagger}(\omega)$, $b(\omega)$ for $-\infty<\omega<\infty$, which obey the usual commutation rules, 
\begin{equation}
[b(\omega),b(\omega ')]=0, \quad [b^{\dagger}(\omega),b^{\dagger}(\omega)']=0, \quad
[b(\omega),b^{\dagger}(\omega ')]=\delta(\omega-\omega ')\nonumber.
\end{equation}
\noindent Then, the operator $W(t)$ (the quantized equivalent of the CSL $w(t)$) and its conjugate 
$\Pi (t)$ may be defined as
\begin{subequations}\label{10} 
\begin{eqnarray}
W(t)&\equiv&(\lambda/2\pi)^{1/2}\int_{-\infty}^{\infty}d\omega [e^{-i\omega t}b(\omega)+
e^{i\omega t}b^{\dagger}(\omega)],\\
\Pi(t)&\equiv&i/(8\pi\lambda)^{1/2}\int_{-\infty}^{\infty}d\omega [-e^{-i\omega t}b(\omega)+
e^{i\omega t}b^{\dagger}(\omega)]
\end{eqnarray}
\end{subequations}
\noindent which are readily shown to satisfy
\begin{equation}\label{11}
[W(t),W(t')]=0, \quad [\Pi(t),\Pi(t')]=0, \quad  [W(t),\Pi(t')]= i\delta(t-t')
\end{equation}
\noindent The vanishing of the first two commutators in Eqs.(11) result from cancellation of 
the positive frequency contribution in Eq.(10) by the negative frequency contribution.  (Incidentally, a free tachyonic 
field also has equal positive and negative energy spectra, so it too can be quantized to have vanishing self-commutator\cite{tachyon}).  
	
	The joint eigenstate of $W(t)$ ($\Pi (t)$) is denoted $|w\rangle$ ($|\pi\rangle$), where the label represents a particular function $w(t)$ ($\pi(t)$) of white noise class.    It satisfies $W(t)|w\rangle=w(t)|w\rangle$ ($\Pi(t)|\pi\rangle=\pi(t)|\pi\rangle$). The range of $t$ is  
$(-\infty,\infty)$, and the states are normalized so that $\int_{-\infty}^{\infty}D'w|w\rangle\langle w|=1$ ($\int_{-\infty}^{\infty}D'\pi|\pi\rangle\langle \pi|=1$).  $D'$ differs from  the $D$  employed for CSL in that it involves the product of $dw(t)$'s for the infinite range of $t$.  To discretize,  write
 $|w\rangle=\prod_{j=-\infty}^{\infty}|w(t_{j}\rangle$, with 
 $ \langle w(t_{j})| w'(t_{j})\rangle=\delta[w(t_{j})-w'(t_{j})]\sqrt{2\pi\lambda/dt}$.

	It follows from Eqs.(11) that, for an arbitrary vector  $|\Phi\rangle$, 
\begin{equation}\label{12}
\langle w|\Pi(t)|\Phi\rangle=\frac{1}{i}\frac{\delta}{\delta w(t)}\langle w|\Phi\rangle.
\end{equation}
\noindent Here, $\delta/\delta w(t)$ is the functional derivative.  In time-discretized calculations, 
$\delta/\delta w(t)\rightarrow (dt)^{-1}\partial /\partial w(t_{j})$.

	The vacuum state $|0\rangle$ satisfies $b(\omega)|0\rangle=0$ as usual.  Since, from Eqs.(10), 
$ W(t)+2i\lambda\Pi(t)$ is an integral over $b(\omega)$,  Eqs.(10), (12) imply  
\begin{equation}
\langle w|W(t)+2i\lambda\Pi(t)|0\rangle=[w(t)+2\lambda\delta/\delta w(t)]\langle w|0\rangle=0\nonumber
\end{equation}
\noindent and a similar equation for $\langle \pi|0\rangle$, whose solutions are 
\begin{equation}\label{13}
\langle w|0\rangle=e^{-(4\lambda)^{-1}\int_{-\infty}^{\infty}dt'w^{2}(t')},\quad 
\langle \pi|0\rangle=e^{-\lambda\int_{-\infty}^{\infty}dt'\pi^{2}(t')}.  
\end{equation}
\noindent The normalization in Eqs.(13) ensure that $\int D'w|\langle w|0\rangle|^{2}=
\int D'\pi |\langle \pi|0\rangle|^{2}=\langle 0|0\rangle=1$.
	
	The $W$-field Hamiltonian, the energy operator which generates time-translations of $W(t)$, $\Pi (t)$, is 
\begin{equation}\label{14}
H_{w}\equiv\int_{-\infty}^{\infty}d\omega \omega b^{\dagger}(\omega)b(\omega)=
\int_{-\infty}^{\infty}dt'\dot{W}(t')\Pi(t'), 
\end{equation}
\noindent which follows from Eqs.(10).  (By adding an infinitesimal imaginary part to $\omega$ 
in Eqs.(10), one can arrange that $W(t)$, $\Pi(t)$ vanish at $t\rightarrow \pm\infty$, so one can freely integrate Eq.(14) by parts, to put the time derivative on $\Pi$.  One can also change the 
order of $\dot{W}(t')$ and $\Pi(t')$ in Eq.(14), since $\int dt'[\dot{W}(t'),\Pi(t')]=0$.)

	The ``ensemble vector" of CQC in the collapse interaction picture is defined as:
\begin{subequations}
\begin{eqnarray}\label{15}
|\psi,t\rangle&\equiv&{\cal T}e^{-2i\lambda\int_{0}^{t}dt''A(t'')\Pi(t'')}|0\rangle|\phi\rangle\\
&=&\int D'w|w\rangle\langle w|
{\cal T}e^{-2i\lambda\int_{0}^{t}dt''A(t'')\Pi(t'')}|0\rangle|\phi\rangle\\ 
&=&\int D'w|w\rangle
{\cal T}e^{-2\lambda\int_{0}^{t}dt''A(t'')(\delta/\delta w(t''))}
e^{-(4\lambda)^{-1}\int_{-\infty}^{\infty}dt'w^{2}(t')}|\phi\rangle\\
&=&\int D'w|w\rangle 
 {\cal T}e^{-(4\lambda)^{-1}\int_{0}^{t}dt'[w(t')-2\lambda A(t')]^{2}}|\phi\rangle
e^{-(4\lambda)^{-1}\{ \int_{-\infty}^{0}dt'w^{2}(t')
+\int_{t}^{\infty}dt'w^{2}(t')\}}
\end{eqnarray}
\end{subequations}

	A way to look at this is as a continuous measurement sequence.  The 
change of the wave function at time $t$ over the interval $dt$, according to (15a),  resides solely in
\[
e^{-2i\lambda A(t)i^{-1}\partial/\partial w(t)}e^{-(4\lambda)^{-1}dtw^{2}(t)}=e^{-(4\lambda)^{-1}dt [w(t)-2\lambda A(t)]^{2}}, 
\]
\noindent which has the form of the von Neumann measurement of $A(t)$ by the pointer $w(t)$. However, here the pointer uncertainty $\sim (dt)^{-1} $ is quite large.  It is only the accumulation of a large number of measurements 	which can distinguish the eigenvalues of $A(t)$.  One can visualize the collection of pointers for all t (and for all $({\bf x}, t)$ in the case of full-blown CSL), each waiting, patiently, 
for its brief opportunity to perform its measurement. 

\subsection{Equivalence of CSL and CQC}

	The time-ordered term in Eq.(15d) is precisely the CSL expression (3b) for 
the state vector which evolves under the classical noise $w(t')$ for $0\leq t'\leq t$.  
The remaining exponentials are unaltered parts of the vacuum state, one for $-\infty< t'<0$ which 
never changes, and  one for $t<t'<\infty$ which evolves.  

	In standard quantum theory, a superselection rule\cite{super} obtains at time $t$ if thereafter the state vector may be written as a 
superposition of orthogonal states, each evolving independently of 
all the others (i.e., the matrix elements of the Hamiltonian between 
these states vanishes).  In such a case, the state vector may be regarded 
as a "catalog," as discussed in the Introduction, merely a convenient device for listing the different states and their probabilities, since any state's evolution after time t can be separately described.

	In CQC, the ensemble vector admits a superselection rule at any time.  Consider the superselected states at time $t$. The labels which distinguish them are the values of $w(t')$ for $0\leq t'\leq t$.  The differently labeled states evolve independently thereafter.  Each state corresponds to a different CSL state.  The values of $w(t')$ for $-\infty<t'< 0$ and  $t< t'\leq \infty$ are irrelevant as far as the particle physics over the interval $(0,t)$ is concerned. The former never becomes relevant. For the latter, as time increases, say from $t$ to $t_{1}$, the values of $w(t')$ for $t<t'< t_{1}$ become relevant: a superselected state at time $t$ splits into many superselected states over the interval $(t,t_{1})$ (in a way that a many-worlds advocate might admire).  Thus, as far as the particle physics is concerned, the ensemble vector may be regarded 
as merely a convenient device for listing the different CSL states, and this is the sense in which CSL and CQC are equivalent.  

	The evolution (15) is unitary and guarantees conservation of probability 
if the usual quantum rule is employed, that the probability of realization of the state $\sim |w\rangle\langle w|\psi,t\rangle$ is $D'w|\langle w|\psi,t\rangle|^{2}$.
 As far as the particle physics is concerned at time $t$, one may integrate the probability of a superselected set over $w(t')$ for 
$t'$ outside the range $0\leq t'\leq t$: the result is the CSL probability $P_{t}(w)Dw$.

\subsection{Full-Blown CQC}

	Full-blown CQC corresponding to full-blown CSL described in Eqs.(8),(9), simply replaces 
time integrals by space-time integrals, e.g., 
\begin{equation}
 W(x)\equiv(\lambda)^{1/2}/(2\pi)^{2}\int d^{4}k [e^{-ik\cdot x}b(k)+
e^{ik\cdot x}b^{\dagger}(k)]\nonumber
\end{equation}
\begin{equation}
|\psi,t\rangle\equiv{\cal T}e^{-2i\lambda\int_{0}^{t}d^{4}x'A(x')\Pi(x')}|0\rangle|\phi\rangle\nonumber
\end{equation}
\noindent replace Eqs.(10a), (15a), etc.

\subsection{Nonwhite Noise CQC}
A CSL collapse evolution which generalizes Eq.(3b)\cite{nonwhite} is
\[
|\psi, t\rangle={\cal T} e^{-(4\lambda)^{-1}\int_{0}^{t}dt'dt''[w(t')-2\lambda A(t')]g(t'-t'')[w(t")-2\lambda A(t")]} |\phi\rangle, 
\]
\noindent where $g(t)$ is real, symmetric in $t$ and has a positive spectrum, i.e., $g(t)=\int d\omega 
\exp (i\omega t)h^{2}(\omega)$ with $h(\omega)=h(-\omega)=h^{*}(\omega)$.  

	A  CQC counterpart requires defining $W(t)$ and $\Pi (t)$ as in Eqs.(10), except that 
the square bracket in the integral of Eq.(10a)  (Eq.(10b)) is multiplied by $h^{-1}(\omega)$ ($h(\omega)$).  This replaces the equation before (13) and Eq.(13a) respectively by 
\[
\big[\int_{-\infty}^{\infty}dt'g(t-t')w(t')+2\lambda\delta/\delta w(t)\big]\langle w|0\rangle=0,\quad
\langle w|0\rangle=e^{-(4\lambda)^{-1}\int_{-\infty}^{\infty}dt'dt''w(t')g(t'-t'')w(t'')}
\]

	With the evolution equation the same as Eq.(15a), Eq.(15d) is replaced by 
\[
|\psi,t\rangle=
 {\cal T}e^{-(4\lambda)^{-1}\int_{-\infty}^{\infty}dt'dt''[w(t')-2\lambda A(t')\Theta_{0,t}(t')]g(t'-t'')[w(t'')-2\lambda A(t'')\Theta_{0,t}(t'')]}|\phi\rangle
\]
\noindent where $\Theta_{0,t}(t')=1$ for $0\leq t'\leq t$ and $ = 0$ otherwise.  For general $h(\omega)$, this equation is {\it not} the same as its CSL counterpart written above, except in the S-matrix limit where the interval $(0,t)$ is replaced by 
$(-\infty, \infty)$.  This generalization shall not be considered any further in this paper.     

\section{Energy}

\subsection{Pictures}
	The CSL Schr\"odinger picture state vector (1) has a corresponding CQC ensemble vector:
\begin{subequations}\label{16}
\begin{eqnarray}
|\psi, t\rangle_{S}&\equiv&e^{-iH_{A}t}|\psi, t\rangle\\
&=&{\cal T} e^{-i\int_{0}^{t}dt'[H_{A}+2\lambda A\Pi(t')]}|0\rangle|\phi\rangle, 
\end{eqnarray}
\end{subequations}
\noindent  where (16b)  follows from putting (15a) into (16a).  Although $A$'s  time dependence in (15a)   is removed in (16b), this is still not the Schr\"odinger  picture for CQC's ensemble vector  (nonetheless, we'll keep the $S$ subscript on it), since the $\Pi$ operator still has time dependence.  It can be removed in what we'll call the sub-Schr\"odinger picture:
\begin{subequations}\label{17}
\begin{eqnarray}
|\psi, t\rangle_{SS}&\equiv&e^{-iH_{w}t}|\psi, t\rangle_{S}\\
&=& e^{-it[H_{w}+H_{A}+2\lambda A\Pi(0)]}|0\rangle|\phi\rangle,
\end{eqnarray}
\end{subequations}

	Thus, since this is a unitary evolution, there is a conserved energy, the Hamiltonian in Eq.(17b). In the collapse-interaction picture it is 	
\[ H(t)=e^{i(H_{A}+H_{W})t}[H_{A}+H_{W}+2\lambda A \Pi (0)]e^{-i(H_{A}+H_{W})t}=
H_{A}+H_{W}+2\lambda A(t) \Pi (t).
\]
	
	It should be emphasized that, since $H_{w}$ and the interaction term depend upon $\Pi$  which has nonvanishing matrix elements between different eigenstates of $W$, the conserved energy belongs to the ensemble of realizable states, not to an individual state. A single collapse to the realized state generally will not preserve the energy distribution. (Of course, this is also the case in standard quantum theory with the collapse postulate.)   Nonetheless, it is interesting to consider the probability distributions of the ensemble total energy and its parts.  A conserved quantity gives insight to the behavior of the ensemble of possibilities.  
	
	For example, one may see how the the ensemble average increase of particle energy, due to the localization mechanism of CSL, is compensated by the ensemble average decrease of the random field energy.  And, laboratory experiments often do involve an ensemble of (approximately) identical initial states, for which therefore energy is (approximately) conserved for the ensemble of outcome states.  Also interesting  are circumstances where one can speak of conservation of energy associated with the ensemble of states that collapses to a distinctly identifiable outcome. For example, when considering a situation when the initial state vector is a superposition of particles in widely separated locations,  the ongoing orthogonality of the differently located particle states ensures energy conservation for the ensemble of states which describe collapse to particles in one of the locations.   
		
\subsection{Simplest Model, $H_{A}=0$}

	We display here generating functions and probability distributions for the total energy and its components, for the model of section IIa, in the simplest case, $H_{A}=0$.  
	
	Using the expression for $|\psi, t\rangle$ given in Eq.(15a),  
the generating function for the moments of the total energy is
\begin{subequations}\label{18}
\begin{eqnarray}
\langle\psi,t| e^{-iH(t)\beta}|\psi, t\rangle&=&\sum_{n}|\alpha_{n}|^{2}
\langle 0| e^{2i\lambda a_{n}\int_{0}^{t}dt' \Pi(t')]}e^{-i[H_{w}+2\lambda a_{n}\Pi(t)]\beta}
e^{-2i\lambda a_{n}\int_{0}^{t}dt' \Pi(t')]}|0\rangle\\
&=&\sum_{n}|\alpha_{n}|^{2}
\langle 0| e^{-i[H_{w}+2i\lambda a_{n}\int_{0}^{t}dt'i\dot{\Pi}(t')+2\lambda a_{n}\Pi(t)]\beta}|0\rangle\\
&=&\sum_{n}|\alpha_{n}|^{2}
\langle 0| e^{-i[H_{w}+2\lambda a_{n}\Pi(0)]\beta}|0\rangle\\
&=&\sum_{n}|\alpha_{n}|^{2}
\langle 0| e^{-iH_{w}\beta}e^{-2i\lambda a_{n}\int_{0}^{\beta}dt'\Pi(t')}|0\rangle\\
&=&\sum_{n}|\alpha_{n}|^{2}\int D' \pi e^{-2\lambda\int_{-\infty}^{\infty}\pi^{2}(t')}
e^{-2i\lambda a_{n}\int_{0}^{\beta}dt'\pi(t')}\\
&=&\sum_{n}|\alpha_{n}|^{2} e^{-(\lambda/2)a_{n}^{2}|\beta |}.
\end{eqnarray}
\end{subequations}	 
\noindent (18b) is the result of performing the unitary transformation on $e^{-iH(t)\beta}$ in (18a), the integration in the exponent of (18b) results in (18c) which confirms that the generating function is time-independent, (18d) expresses the unitary operator in (18c) in a different form, (18e) follows from $H_{w}|0\rangle=0$ and insertion of the identity $1=\int D'\pi |\pi\rangle\langle \pi |$, (18f) follows by discretization and performance of the gaussian integrals.

	The moments of the energy follow from Eq.(18f) using the formal Taylor series expansion, 
$|\beta|=0+\beta\epsilon (0)+(\beta^{2}/2!)2\delta(0)+(\beta^{3}/3!)2\delta '(0)+...$  
(where $\epsilon (x)\equiv x/|x|$ for $x\neq 0$ and $\epsilon (0)\equiv 0$).  
Thus the mean energy is 0 and the higher moments of the energy are infinite.  The associated probability distribution more transparently entails this result:
\begin{subequations}\label{19}
\begin{eqnarray}
{\cal P}(E)&\equiv&\frac{1}{2\pi}\int_{-\infty}^{\infty}e^{iE\beta}\langle\psi,t| e^{-iH(t)\beta}|\psi, t\rangle\\
&=&\frac{1}{\pi}\sum_{n}|\alpha_{n}|^{2}\frac{\lambda a_{n}^{2}/2}{E^{2}+(\lambda a_{n}^{2}/2)^{2}}.
\end{eqnarray}
\end{subequations}	 

	Although the total energy probability distribution is time-independent, the 
energy $H_{w}$ is not.  The generating function here is 
\begin{subequations}\label{20}
\begin{eqnarray}
\langle\psi,t| e^{-iH_{w}\beta}|\psi, t\rangle&=&\sum_{n}|\alpha_{n}|^{2}
\langle 0| e^{2i\lambda a_{n}\int_{0}^{t}dt' \Pi(t')}e^{-2i\lambda a_{n}\int_{0}^{t}dt' \Pi(t'-\beta)}|0\rangle\\
&=&\sum_{n}|\alpha_{n}|^{2}\int D' \pi e^{-2\lambda\int_{-\infty}^{\infty}\pi^{2}(t')}
e^{2i\lambda a_{n}[\int_{0}^{t} dt-\int_{-\beta}^{t-\beta}]dt'\pi(t')}\\
&=&\sum_{n}|\alpha_{n}|^{2}\big\{[\Theta(\beta-t) + \Theta (-t-\beta)]e^{-\lambda a_{n}^{2}t} \nonumber\\
&&\qquad\qquad\qquad\qquad+\Theta (t-\beta)\Theta (t+\beta)e^{-\lambda a_{n}^{2}|\beta |}\big\} .
\end{eqnarray}
\end{subequations}	 	 
\noindent where (20a) utilizes (15a) and the time-translation capability of $H_{w}$, (20b) involves going to the $|\pi\rangle$ representation, and (20c) follows from performing the gaussian integrals.  From the Taylor series expansion of (20c), one sees that the mean value of $H_{w}$ is zero and the higher moments are infinite.  

	The probability distribution following from Eq.(20c) confirms this result:  
 \begin{subequations}\label{21}
\begin{eqnarray}
{\cal P}_{w}(E)&\equiv&\frac{1}{2\pi}\int_{-\infty}^{\infty}e^{iE\beta}\langle\psi,t| e^{-iH_{w}\beta}|\psi, t\rangle\\
&=&\frac{1}{\pi}\sum_{n}|\alpha_{n}|^{2}\bigg \{e^{-\lambda a_{n}^{2}t}\bigg [\pi\delta(E) -
\frac{\sin Et}{E}+\frac{E\sin Et-\lambda a_{n}^{2}\cos Et}{E^{2}+(\lambda a_{n}^{2})^{2}}\bigg ] \nonumber \\
&&\qquad\qquad\qquad\qquad\qquad\qquad\qquad\qquad
+\frac{\lambda a_{n}^{2}}{E^{2}+(\lambda a_{n}^{2})^{2}}\bigg \}. 
\end{eqnarray}
\end{subequations}	
\noindent Eq. (21) shows how the $W$-field energy, starting at $t=0$ with the vacuum energy $E= 0$, evolves toward a stationary energy distribution associated with the completely collapsed state.  (Note that the asymptotic value of (21b) is {\it not} the same as (19b), although it is similar in form.)

	The interaction energy's probability distribution is 
 \begin{subequations}\label{22}
\begin{eqnarray}
{\cal P}_{I}(E)&\equiv&\frac{1}{2\pi}\int_{-\infty}^{\infty}e^{iE\beta}\langle\psi,t| e^{-iH_{I}\beta}|\psi, t\rangle\\
&=&\frac{1}{2\pi}\int_{-\infty}^{\infty}e^{iE\beta}
\sum_{n}|\alpha_{n}|^{2}\int D' \pi e^{-2\lambda\int_{-\infty}^{\infty}dt'\pi^{2}(t')}
e^{-2i\lambda a_{n}\pi(t)\beta}\\
&=&\sum_{n}\frac{|\alpha_{n}|^{2}}{\sqrt{2\pi \lambda a_{n}^{2}/dt}}e^{-E^{2}/(2\lambda a_{n}^{2}/dt)}
\end{eqnarray}
\end{subequations}
\noindent Since the interaction energy is proportional to the white noise operator $\Pi(t)$, it has a white noise energy spectrum, with all energies equally likely.   
 		 
\subsection{Simplest Model, $H_{A}\neq 0$}
\subsubsection{Moment Generating Functions}
	As in the previous section, also in this more general case, the random field variables can be integrated out of the expressions for the moment generating functions.  For the total energy, corresponding to Eq.(18), one obtains:
\begin{subequations}\label{23}
\begin{eqnarray}
\langle\psi,t| e^{-iH(t)\beta}|\psi, t\rangle&=&
\langle \phi|\langle 0| e^{-i[H_{A}+H_{w}+2\lambda\Pi(0)A]\beta}|0\rangle|\phi\rangle\\
&=&\langle \phi| \langle 0|e^{-i[H_{A}+H_{w}]\beta}{\cal T}e^{-2i\lambda \int_{0}^{\beta}dt'\Pi(t')A(t')}|0\rangle|\phi\rangle\\
&=&\langle \phi| e^{-iH_{A}\beta} \int D'\pi e^{-2\lambda\int_{-\infty}^{\infty}dt'\pi^{2}(t')}{\cal T}e^{-2i\lambda \int_{0}^{\beta}dt'\pi(t')A(t')}|\phi\rangle\\
&=&\langle \phi| e^{-iH_{A}\beta} {\cal T}e^{-(\lambda/2) \int_{0}^{\beta}dt'A^{2}(t')}|\phi\rangle\\
&=&\langle \phi| e^{-[iH_{A} +(\lambda/2)A^{2}(0)]\beta}|\phi\rangle. 
\end{eqnarray}
\end{subequations}
\noindent ((23a) is expressed in the manifestly time-independent sub-Schr\"odinger picture, and the remaining steps are ones taken before.)
	For $H_{A}$, the moment generating function is 
\begin{subequations}\label{24}
\begin{eqnarray}
\langle\psi,t| e^{-iH_{A}\beta}|\psi, t\rangle&=&	 	 
\langle \phi|\langle 0|{\cal T}e^{2i\lambda \int_{0}^{t}dt'\Pi(t')A_{L}(t')} e^{-iH_{A}\beta}
e^{-2i\lambda \int_{0}^{t}dt'\Pi(t')A_{R}(t')}|0\rangle|\phi\rangle\\
&=&\langle\phi |{\cal T}\big\{e^{ -(\lambda/2)\int_{0}^{t}dt'[A_{L}(t')-A_{R}(t')]^{2}}e^{-iH_{A}\beta}\big\}|\phi\rangle,
\end{eqnarray}
\end{subequations}
\noindent where $A_{R}$ ($A_{L}$) acts to the right (left) of $\exp-iH_{A}\beta$, and is time-ordered (time--reverse-ordered) by ${\cal T}$.  

	For $H_{w}$, the calculation is  
\begin{subequations}\label{25}
\begin{eqnarray}
\langle\psi,t| e^{-iH_{w}\beta}|\psi, t\rangle&&=	 	 
\langle \phi|\langle 0|{\cal T}e^{2i\lambda \int_{0}^{t}dt'\Pi(t')A_{L}(t')} e^{-iH_{w}\beta}
e^{-2i\lambda \int_{0}^{t}dt'\Pi(t')A_{R}(t')}|0\rangle|\phi\rangle\\
&&=\langle \phi|\langle 0|{\cal T}e^{2i\lambda \int_{0}^{t}dt'\Pi(t'+\beta/2)A_{L}(t')}e^{-2i\lambda \int_{0}^{t}dt'\Pi(t'-\beta/2)A_{R}(t')}|0\rangle|\phi\rangle\\
&&=\langle \phi|\int D' \pi e^{-2\lambda\int_{-\infty}^{\infty}dt'\pi^{2}(t')}     \notag      \\ 
&&\cdot {\cal T}e^{2i\lambda \int_{\beta/2}^{t+\beta/2}dt'\pi(t')A_{L}(t'-\beta/2)}e^{-2i\lambda\int_{-\beta/2}^{t-\beta/2}dt'\pi(t')A_{R}(t'+\beta/2)}|\phi\rangle\\
&&=\langle \phi|{\cal T}e^{ -(\lambda/2)\int_{0}^{t}dt'[A_{L}^{2}(t')+A_{R}^{2}(t')]}
\big[\Theta (\beta -t)+\Theta (-\beta -t) \nonumber\\
&&\qquad+\Theta (t-\beta)\Theta (t+\beta)e^{\lambda
\int_{|\beta | /2}^{t-|\beta | /2}dt' A_{L}(t'-\beta/2)A_{R}(t'+\beta/2)}\big]|\phi\rangle
\end{eqnarray}
\end{subequations}	 

 \subsubsection{First Moments}
 
 	Here is shown conservation of energy in the mean, reproducing more simply a previously obtained result (all that that had been proved about conservation of energy for CSL\cite{energy}).   
 	Equating the terms proportional to the first power of $\beta$ in Eq,(24b) produces	
\begin{equation}\label{26}
 \langle\psi,t| H_{A}|\psi, t\rangle=	 	 
\langle\phi |{\cal T}e^{ -(\lambda/2)\int_{0}^{t}dt'[A_{L}(t')-A_{R}(t')]^{2}}H_{A}|\phi\rangle.
\end{equation}		
	To find the first moment of $H_{w}$,  the exponent in (25d) must be written for small $\beta$:
\begin{eqnarray}\label{27}
&&\int_{|\beta | /2}^{t-|\beta | /2}dt' A_{L}(t'-\beta/2)A_{R}(t'+\beta/2)=
\int_{0}^{t}dt' A_{L}(t')A_{R}(t')\nonumber\\
&&\negthinspace\negthinspace\negthinspace\negthinspace\negthinspace\negthinspace\negthinspace\negthinspace\negthinspace\negthinspace
-(|\beta| /2)[A_{L}(t)A_{R}(t)+A_{L}(0)A_{R}(0)]
+(\beta /2)\int_{0}^{t}dt' [A_{L}(t')\dot{A}_{R}(t')- \dot{A}_{L}(t')A_{R}(t')]+...
\end{eqnarray}
 \noindent The zeroth and first order terms in the Taylor series expansion of (27) are the same as (27),  except that the second term vanishes, since $d|\beta|/dt|_{\beta=0}=\epsilon(0)=0$. The step-functions in (25d) likewise do not contribute to the Taylor series expansion's first two terms, so we get from (25d):
\begin{subequations}\label{28}
\begin{eqnarray}
 \langle\psi,t| H_{w}|\psi, t\rangle&=&i(\lambda/2)  	 	 
\langle\phi |{\cal T}e^{ -(\lambda/2)\int_{0}^{t}dt'[A_{L}(t')-A_{R}(t')]^{2}}\nonumber\\ 
&&\qquad\qquad\cdot \int_{0}^{t}dt' [ A_{L}(t')\dot{A}_{R}(t')- \dot{A}_{L}(t')A_{R}(t')] |\phi\rangle\\
&=&(\lambda/2)\langle\phi |{\cal T}\int_{0}^{t}dt' e^{ -(\lambda/2)\int_{0}^{t'}dt''[A_{L}(t'')-A_{R}(t'')]^{2}}[A(t'),i\dot{A}(t')]
 |\phi\rangle\\
&=&(\lambda/2)\langle\phi |{\cal T}\int_{0}^{t}dt' e^{ -(\lambda/2)\int_{0}^{t'}dt''[A_{L}(t'')-A_{R}(t'')]^{2}}[A(t'),[{A}(t'),H_{A}]] |\phi\rangle
\end{eqnarray}
\end{subequations}	   
 \noindent In going from (28a) to (28b) we have utilized  
\[ 
{\cal T}e^{ -(\lambda/2)\int_{0}^{t}dt'[A_{L}(t')-A_{R}(t')]^{2}}B_{L}(t_{1})C_{R}(t_{1})=
{\cal T}e^{ -(\lambda/2)\int_{0}^{t_{1}}dt'[A_{L}(t')-A_{R}(t')]^{2}}B(t_{1})C(t_{1})
\]
\noindent for $t_{1}<t$ and arbitrary operators $B_{L}(t_{1})$, $C_{R}(t_{1})$ (the subscripts $R$, $L$ may be dispensed with when operators having the same time argument are adjacent and in the correct order).  

	Since the first moment of the 
interaction energy vanishes, from Eqs.(26), (28c) follows conservation of energy in the mean:
\begin{eqnarray}\label{29}
(d/dt) \langle\psi,t| H_{w}|\psi, t\rangle&=&-(d/dt)\langle\psi,t| H_{A}|\psi, t\rangle\nonumber\\
&=&(\lambda/2)\langle\phi |{\cal T} e^{ -(\lambda/2)
\int_{0}^{t}dt'[A_{L}(t')-A_{R}(t')]^{2}}[A(t),[{A}(t),H_{A}]] |\phi\rangle
\end{eqnarray}

\subsubsection{Large t}

	The large t-behavior of the probability density distribution of $H_{w}$ is due to the 
third term in the bracket of Eq.(25d).  Because this term's exponent is largest for small $\beta$, and also because $\exp iE\beta$ oscillates rapidly for large $\beta$, the small $\beta$ approximation of (25d) should be a good approximation to the large t limit. Putting (27) into (25d) results in
\begin{subequations}\label{30}
\begin{eqnarray}
&&{\cal P}_{w}(E)=\frac{1}{2\pi}\int_{-\infty}^{\infty}e^{iE\beta}\langle\psi,t| e^{-iH_{w}\beta}|\psi, t\rangle\\
&&\negthinspace\negthinspace\negthinspace\negthinspace\negthinspace\negthinspace\negthinspace\negthinspace\negthinspace\negthinspace\negthinspace\negthinspace\negthinspace\negthinspace
\approx\frac{1}{2\pi}\int_{-t}^{t}d\beta e^{iE\beta}\langle\phi|{\cal T} 
e^{ -(\lambda/2)\int_{0}^{t}dt'[A_{L}(t')-A_{R}(t')]^{2}}e^{-(\lambda /2)|\beta|[A^{2}(t)+A^{2}(0)]}
e^{-i(\lambda /2)\beta\int_{0}^{t}dt'C_{LR}(t')}|\phi\rangle\\
&&\negthinspace\negthinspace\negthinspace\negthinspace\negthinspace\negthinspace\negthinspace\negthinspace\negthinspace\negthinspace\negthinspace\negthinspace\negthinspace\negthinspace
=\langle\phi|{\cal T} 
e^{ -(\lambda/2)\int_{0}^{t}dt'[A_{L}(t')-A_{R}(t')]^{2}} 
\frac{1}{\pi}
\frac{\lambda [A^{2}(t)+A^{2}(0)]/2}{[E-(\lambda /2)\int_{0}^{t}dt'C_{LR}(t')]^{2}+[\lambda (A^{2}(t)+A^{2}(0))/2]^{2}}|\phi\rangle
 \nonumber\\
\end{eqnarray}
\end{subequations}	 	 
\noindent where $C_{LR}(t')\equiv A_{L}(t')[A_{R}(t'),H_{A}]-[A_{L}(t'),H_{A}]A_{R}(t')$.  Note that,  when $H_{A}=0$,
the large t-limit of (30c)  is the same as (21b). Also, note  that, when $H_{A}\neq0$,
 the mean value of $E$ given by (30c) is the same as (28c).
 
 	For example, in the case of a free particle of mass $m$ which collapses toward a position eigenstate,  where $H=p^{2}/2m$ and $A=x$, the term in (30c) $[E-(\lambda /2)\int_{0}^{t}dt'C_{LR}(t')]^{2}=
[E+\lambda\hbar t/2m]^{2}$.  Because the ensemble energy is conserved, the mean $W$-field energy decreases with time to compensate for the  increase of the particle's mean energy $=\lambda\hbar t/2m$ due to the narrowing of wavepackets by the collapse mechanism.  
 
	\section{Time Operator}
	
	Unlike the usual Hamiltonian which is semi-bounded\cite{timeref},  $H_{w}$ has an unbounded spectrum, providing an opportunity to define a self-adjoint  time operator:
	\begin{subequations}\label{31}
\begin{eqnarray}
T&\equiv&\frac{ \int_{-\infty}^{\infty}d\omega b^{\dagger}(\omega)\frac{1}{i}\frac{\partial}{\partial \omega}b(\omega)}{
 \int_{-\infty}^{\infty}d\omega b^{\dagger}(\omega)b(\omega)}\equiv \frac{B}{N}\\
 &=&\frac{ \int_{-\infty}^{\infty}dt' t'[(4\lambda)^{-1}W^{2}(t')+\lambda \Pi^{2}(t')]} 
 { \int_{-\infty}^{\infty}dt' [(4\lambda)^{-1}W^{2}(t')+\lambda \Pi^{2}(t')]},
 \end{eqnarray}
\end{subequations}	
\noindent where $[B,N]=0$.  It is simple to show that $[H_{w},T]=i$, where $H_{w}$ is given by Eq.(14).  The action of $T$ on $|0\rangle$, is undefined, but one can
take $T=B/(N+\epsilon)$ so that $T |0\rangle=0/\epsilon =0$. Of course, $T$ plus any function of $H_{w}$ and $N$ is also 
conjugate to $H_{w}$, but (31)  is the simplest  choice.

	\subsection{Eigenfunctions} 
	
	$T$'s eigenvalues are highly degenerate.  A complete set of (unnormalized) eigenfunctions in the $|w\rangle$ basis is:
\begin{eqnarray}
\langle w| \prod_{t}[\int_{-\infty}^{\infty}d\omega e^{i\omega t}b^{\dagger}(\omega)]^{n(t)}|0\rangle
&\sim& \prod_{t}[ w(t)-2\lambda\delta/\delta\omega ]^{n(t)}e^{-(4\lambda)^{-1}\int_{-\infty}^{\infty}dt' w^{2}(t')}
\nonumber\\
&\sim& \prod_{t}H_{n(t)}[w(t)(dt/2\lambda)^{1/2}]e^{-(4\lambda)^{-1}\int_{-\infty}^{\infty}dt' w^{2}(t')}\nonumber
\end{eqnarray}
\noindent where $n(t)=0,1,2...$ and the $H_{n}$ are Hermite polynomials.  The associated 
eigenvalues $\sum_{t=-\infty}^{\infty}t n(t)/\sum_{t=-\infty}^{\infty}n(t)$ have a ``center of time" form.

\subsection{Moment Generating Function}

	To find the moment generating function for $T$, first observe, using (15a) and (10b), that 

\begin{eqnarray}
 &&e^{-i\beta T}|\psi, t\rangle=e^{-i\beta T}{\cal T}e^{(\lambda/2\pi)^{1/2}\int_{0}^{t}dt'
 A_{R}(t')\int _{-\infty}^{\infty}d\omega e^{i\omega t'} b^{\dagger}(\omega)}|0\rangle 
 e^{-(\lambda/2)\int_{0}^{t}dt' A^{2}_{R}(t')}|\phi\rangle\nonumber \\
 &&\quad= {\cal T}\bigg[1+\sum_{n=1}^{\infty}e^{-i\beta B/n}\frac{(\lambda/2\pi)^{n/2}}{n!}
 \big[\int_{0}^{t}dt' A_{R}(t')\int _{-\infty}^{\infty}d\omega e^{i\omega t'} b^{\dagger}(\omega)\big]^{n}\bigg]
 |0\rangle 
 e^{-(\lambda/2)\int_{0}^{t}dt' A^{2}_{R}(t')}|\phi\rangle. \nonumber 
\end{eqnarray}	
\noindent Since $\exp -(i\beta B/n) b^{\dagger}(\omega)\exp (i\beta B/n)= b^{\dagger}
(\omega+\beta/n)$, one gets:
\begin{subequations}\label{32}
\begin{eqnarray}
\langle\psi,t | e^{-i\beta T}|\psi, t\rangle&=&\langle\phi | {\cal T} 
e^{-(\lambda/2)\int_{0}^{t}dt' [A^{2}_{R}(t')+A^{2}_{L}(t')]}\langle 0| 
\bigg[1+\sum_{n=1}^{\infty}\frac{(\lambda/2\pi)^{n/2}}{n!}
 \big[\int_{0}^{t}dt' A_{L}(t')\int _{-\infty}^{\infty}d\omega e^{-i\omega t'} b(\omega)\big]^{n}\bigg]\nonumber\\
&&\quad\cdot\bigg[1+\sum_{n=1}^{\infty}\frac{(\lambda/2\pi)^{n/2}}{n!}
 \big[\int_{0}^{t}dt' A_{R}(t')\int _{-\infty}^{\infty}d\omega e^{i(\omega-n^{-1}\beta )t'} b^{\dagger}(\omega)\big]^{n}\bigg]|0\rangle|\phi\rangle\\
&=&\langle\phi | {\cal T} 
e^{-(\lambda/2)\int_{0}^{t}dt' [A^{2}_{R}(t')+A^{2}_{L}(t')]}
\bigg[1+\sum_{n=1}^{\infty}\frac{(\lambda)^{n}}{n!}
 \big[\int_{0}^{t}dt' A_{L}(t')A_{R}(t')e^{-i\beta t'/n}\big]^{n}\bigg]|\phi\rangle
\end{eqnarray}
\end{subequations}	
\noindent	 As a check, when $\beta=0$, (32b) becomes:
\[
\langle\phi | {\cal T} 
e^{-(\lambda/2)\int_{0}^{t}dt' [A^{2}_{R}(t')+A^{2}_{L}(t')]}
e^{\lambda\int_{0}^{t}dt' [A_{R}(t')A_{L}(t')]}|\phi\rangle=
\langle\phi | {\cal T} 
e^{-(\lambda/2)\int_{0}^{t}dt' [A_{R}(t')-A_{L}(t')]^{2}}
|\phi\rangle=1.
\]

\subsection{First Moments}
	Now, extract the first two moments of $T$ from (32b).  From the coefficient of $\beta$ in (32b) we obtain:
\begin{eqnarray}\label{33}
\langle\psi,t | T|\psi, t\rangle&=&\langle\phi | {\cal T} \frac{\int_{0}^{t}dt' t'A_{L}(t')A_{R}(t')}
{\int_{0}^{t}dt' A_{L}(t')A_{R}(t')}\nonumber\\
&&\qquad\cdot\big[e^{-(\lambda/2)\int_{0}^{t}dt' [A_{L}(t')-A_{R}(t')]^{2}}
-e^{-(\lambda/2)\int_{0}^{t}dt' [A^{2}_{L}(t')+A^{2}_{R}(t')]}\big]|\phi\rangle
\end{eqnarray}
	
	To get a feel for the meaning of (33), consider the case $H_{A}=0$:
\begin{eqnarray}\label{34}
\langle\psi,t | T|\psi, t\rangle=\sum_{k}|\alpha_{k}|^{2}\frac{\int_{0}^{t}dt't'}{\int_{0}^{t}dt}
[1-e^{-\lambda t a_{k}^{2}}]=
\frac{t}{2}[1-\sum_{k}|\alpha_{k}|^{2}e^{-\lambda t a_{k}^{2}}].
\end{eqnarray}
\noindent Thus, asymptotically, $ \langle\psi,t | T|\psi, t\rangle$ approaches 
the ``center of time," a weighted average (here uniformly weighted) of the 
time over the interval $(0,t)$.  For $H_{A}\neq 0$ much the same behavior prevails, according to Eq.(33).  For example, set $A_{R}(t')=A_{L}(t')=At'^{s}$.  Then, from (33), asymptotically, 
$ \langle\psi,t | T|\psi, t\rangle\rightarrow t(2s+1)/(2s+2)$: for $s$ large, the mean value of $T$ 
is almost $t$.

	For full-blown CQC, the expression that replaces (33) is  the same expression  
with $A(t')\rightarrow A({\bf x},t')$ and $dt'\rightarrow d{\bf x}dt'$.   $A$ (given by 
Eq. (8)) is $\approx$ the mass density in a sphere of radius $a$.  Then the appropriate 
``center of time " expression appearing in the equivalent of (33) is 
\[ 
\int_{0}^{t}dt' d{\bf x}t'A_{L}({\bf x},t')A_{R}({\bf x},t')
/\int_{0}^{t}dt' d{\bf x}A_{L}({\bf x},t')A_{R}({\bf x},t'), 
\]
\noindent i.e., $t'$, roughly speaking,  is weighted by 
the squared mass density integrated over all space and over time from 0 to $t$.  Should there 
be a high rate of agglomeration of matter, e.g., due to gravitational clumping, 
then $ \langle\psi,t | T|\psi, t\rangle$ will approach $t$.  It should be emphasized that gravitational 
collapse has this effect, not wave function collapse, as Eq. (34) attests.  

	The second moment of $T$ is likewise found from (32b):	

\begin{eqnarray}\label{35}
 \langle\psi,t | T^{2}|\psi, t\rangle&=&\langle\phi | {\cal T}\bigg\{\bigg[ \frac{\int_{0}^{t}dt' t'A_{L}(t')A_{R}(t')}
{\int_{0}^{t}dt' A_{L}(t')A_{R}(t')}\bigg]^{2}\nonumber\\
&&\quad\cdot\big[ e^{-(\lambda/2)\int_{0}^{t}dt' [A_{L}(t')-A_{R}(t')]^{2}}-
e^{-(\lambda/2)\int_{0}^{t}dt' [A^{2}_{L}(t')+A^{2}_{R}(t')]}[1+f(Z_{LR})]\big]\nonumber\\
&&\quad+ \frac{\int_{0}^{t}dt' t'^{2}A_{L}(t')A_{R}(t')}
{\int_{0}^{t}dt' A_{L}(t')A_{R}(t')}e^{-(\lambda/2)\int_{0}^{t}dt' [A^{2}_{R}(t')+A^{2}_{L}(t')]}f(Z_{LR})\bigg\}|\phi\rangle,
\end{eqnarray}
\noindent where $Z_{LR}\equiv \lambda\int_{0}^{t}dt' A_{L}(t')A_{R}(t')$ and
 $f(z)\equiv\int_{0}^{z}dz'[\exp z' -1]/z'$.	
\noindent For the example $A_{R}(t')=A_{L}(t')=At'^{s}$ ($H_{A}=0$ corresponds to $s=0$), Eq.(35) asymptotically becomes (using $f(z)\rightarrow z^{-1}\exp z$):
\begin{equation}\label{36}
\langle\psi,t | T^{2}|\psi, t\rangle\rightarrow\bigg[\frac{t(2s+1)}{2s+2}\bigg]^{2}
\bigg[1+\frac{1}{\lambda t^{2s+1}}\frac{(2s+2)^{2}}{2s+3}\sum_{k}\frac{|\alpha_{k}|^{2}}{a_{k}^{2}}\bigg].
\end{equation}
\noindent  so the fractional deviation decreases over time, $\overline{\Delta T^{2}}/\overline{T}^{2}\sim t^{-(2s+1)}$.  
 	
\subsection{Large t}

	More generally, an approximation to the large $t$ probability distribution of $T$ may be had 
by employing the small $\beta$ approximation in  (32b),
\begin{eqnarray}
[\int_{0}^{t}dt' A_{L}(t')A_{R}(t')e^{-i\beta t'/n}]^{n}&\approx &[\int_{0}^{t}dt' A_{L}(t')A_{R}(t')(1-i\beta t'/n)]^{n}\nonumber\\
&\approx &[\int_{0}^{t}dt' A_{L}(t')A_{R}(t')]^{n}e^{-i\beta\int_{0}^{t}dt't' A_{L}(t')A_{R}(t')/\int_{0}^{t}dt' A_{L}(t')A_{R}(t')},\nonumber
\end{eqnarray}
\noindent with the result
\begin{eqnarray}
{\cal P}(\tau)&\equiv &(2\pi)^{-1}\int_{-\infty}^{\infty}d\beta e^{i\beta \tau}\langle\psi,t | e^{-i\beta \tau}|\psi, t\rangle\nonumber\\
&\approx&\langle\phi | {\cal T}e^{-(\lambda/2)\int_{0}^{t}dt' [A_{L}(t')-A_{R}(t')]^{2}}
\delta\bigg[ \tau- \frac{ \int_{0}^{t}dt't' A_{L}(t')A_{R}(t')}{\int_{0}^{t}dt' A_{L}(t')A_{R}(t')}\bigg]|\phi\rangle.  
\end{eqnarray}

	To summarize, there is a natural clock associated with the random  field, 
whose time reading depends upon global dynamics and whose deviance decreases with time.  In future work, rather than collapse depending upon time, one may look to express the attractive idea that the notion of time may depend upon collapse, since no collapse implies no events implies no time.  
	
\section{Two examples}
	Collapse models are phenomenological, a guess (based upon a perceived flaw in standard quantum theory) that nature utilizes this description 
of dynamical behavior.  To be generally accepted, such models need 
experimental verification\cite{exp} and
grounding within a larger physical framework.  One of the main aims of this paper has been to emphasize that CSL can be formulated  as CQC, within an essentially quantum framework, in order to encourage thoughts about how collapse might arise from a physical structure which has a natural quantum theory expression.  In that spirit,  
this section contains two examples of CQC dynamics arising from other quantum conceptual structures.

 \subsection{Quantum Fields}
 
	 Consider a collection of scalar quantum fields, associated with particles of large mass $M$  
which, for definiteness, take to be  the Planck mass, so large that 
$\sqrt{{\bf k}^{2}+M^{2}}\approx M$ for any so-far experimentally achievable ${\bf k}$.  Each field  (more precisely, its time derivative, which is its conjugate field) interacts with the mass density $A({\bf x},t)$  (given in Eq.(8) et. seq.) over a brief interval which, for definiteness, we shall imagine is the Planck time $\tau=M^{-1}$.  Call the field $\Phi_{t}({\bf x})$ (and call its conjugate $\Pi_{t}({\bf x})$) which is  responsible for the interaction over the interval centered at time $t$;
\begin{subequations}\label{38}
\begin{eqnarray}
\Phi_{t}({\bf x})&\equiv& C\frac{1}{(2\pi)^{3/2}\sqrt{2M}}\int {\bf dk}[e^{i({\bf k}\cdot{\bf x}-Mt)} 
b_{t}({\bf k})+e^{-i({\bf k}\cdot{\bf x}-Mt)} b_{t}^{\dagger}({\bf k})]\\
\Pi_{t}({\bf x})&\equiv& C^{-1}\frac{i}{(2\pi)^{3/2}}\sqrt{\frac{{M}}{{2}}}\int {\bf dk}[-e^{i({\bf k}\cdot{\bf x}-Mt)} 
b_{t}({\bf k})+e^{-i({\bf k}\cdot{\bf x}-Mt)} b_{t}^{\dagger}({\bf k})]
\end{eqnarray}
\end{subequations}	
\noindent where the constant $C$ is yet to be chosen.  

	Define the eigenstates by $\Phi_{t}({\bf x})|\varphi_{t}\rangle 
=\varphi_{t}({\bf x})|\varphi_{t}\rangle $ and the associated vacuum state by $b_{t}({\bf k})|0\rangle_{t}=0$. Then, Eqs.(38) imply 
\begin{equation}\label{39}
\big[\frac{M}{C^{2}}\varphi_{t}({\bf x})+\frac{\delta}{\delta \varphi_{t}({\bf x})}\big]\langle \varphi_{t}({\bf x})|0\rangle_{t}=0, \quad\langle\varphi_{t}({\bf x})|0\rangle_{t}=e^{-\frac{M}{2C^{2}}\int d{\bf x}\varphi_{t}({\bf x})^{2}}.
\end{equation}
\noindent The joint vacuum state for all the fields is $|0\rangle\equiv\prod_{t}|0\rangle_{t}$ and a joint eigenstate of all the fields is $|\varphi\rangle\equiv\prod_{t}|\varphi_{t}({\bf x})\rangle_{t}$. To make their scalar product close to  the scalar product  in Eq.(13) (suitably generalized to space-time),  choose $C=M\sqrt{2\lambda}$ so, from (39), 
\begin{equation}\label{40}
\langle\varphi |0\rangle=e^{-(4\lambda)^{-1}\sum_{t}\tau\int d{\bf x}\varphi_{t}({\bf x})^{2}}
\approx e^{-(4\lambda)^{-1}\int_{-\infty}^{\infty}dt\int d{\bf x}\varphi_{t}({\bf x})^{2}}
\end{equation}

	The evolution equation is chosen as:  
\begin{subequations}\label{41}
\begin{eqnarray}
\negthinspace\negthinspace\negthinspace\negthinspace\negthinspace\negthinspace
\negthinspace\negthinspace\negthinspace\negthinspace\negthinspace\negthinspace
\langle\varphi | \psi,t\rangle&\equiv&{\cal T}
e^{-2i\lambda\sum_{t'=0}^{t}\int d{\bf x}'\Pi_{t'}({\bf x})A({\bf x}', t')}|0\rangle|\phi\rangle \approx
e^{-2i\lambda\int_{t'=0}^{t}dt'\int d{\bf x}'M\Pi_{t'}({\bf x}')A({\bf x}', t')}|0\rangle|\phi\rangle\\
&=&{\cal T}e^{-(4\lambda)^{-1}\sum_{t'=0}^{t}\tau \int d{\bf x}'[\varphi_{t'}({\bf x}')-2\lambda A({\bf x}',t')]^{2}}|\phi\rangle
e^{-(4\lambda)^{-1}[ \sum_{t'=-\infty}^{0}+ \sum_{t'=t}^{\infty}] \tau \int d({\bf x}')\varphi_{t'}^{2}({\bf x}')}
\end{eqnarray}
\end{subequations}	

	To complete the demonstration that this model is equivalent to CQC,  identify the fields
\begin{equation}\label{42}
\tilde{W}({\bf x},t)\equiv \Phi_{t}({\bf x}),\quad \tilde{\Pi}({\bf x},t)\equiv M \Pi_{t}({\bf x})
\end{equation}
\noindent from which follows 
\[
[\tilde{W}({\bf x},t), \tilde{\Pi}({\bf x}',t')]=\delta({\bf x}-{\bf x}')\delta_{t,t'}/\tau\approx \delta({\bf x}-{\bf x}')\delta(t-t').
\]
\noindent It is immediately seen, using Eq.(42) for  $\tilde{\Pi}({\bf x},t)$, that the statevector evolution equation (41a) is identical to Eq.(15a) (generalized to space-time).

	To see that the expressions for $\tilde{W}$ and $\tilde{\Pi}$ in terms of particle creation and annihilation operators closely approximate the similar expressions for $W$ and $\Pi$,  note that 
\begin{equation}\label{43}
[e^{-iMt}b_{t}({\bf k})/\sqrt{\tau}, e^{iMt}b_{t}^{\dagger}({\bf k})/\sqrt{\tau}]=\delta({\bf k}-{\bf k}')
\delta_{t,t'}/\tau\approx\delta({\bf k}-{\bf k}')\delta(t-t').
\end{equation}
\noindent Therefore,  define 
\begin{equation}\label{44}
\tilde{b}({\bf k}, \omega)\equiv(2\pi)^{-1/2}\int dt' e^{i\omega t'}e^{-iMt'}b_{t'}({\bf k}')/\sqrt\tau, 
\end{equation}
\noindent so $[\tilde{b}({\bf k}, \omega),\tilde{b}^{\dagger}({\bf k}', \omega ']=\delta({\bf k}-{\bf k}')\delta(\omega-\omega ')$  follows from Eqs.(43),(44).  Inserting the inverse of Eq.(44), 
for $b_{t}({\bf k})$ in terms of $\tilde{b}({\bf k}, \omega)$, into Eqs.(38),(42) gives for  $\tilde{W}$ and $\tilde{\Pi}$ expressions whose form is identical to (10)'s expressions (generalized to space-time) for $W$ and $\Pi$.  

\subsection{Many Spins} 
	 
	The model described here borrows from a gravitationally based 
semi-classical collapse model\cite{pearlesquires} and spin-network models of 
space-time\cite{smolinseth}, without being faithful to either.  Suppose that space consists 
of ``Planck cells" (each of volume $\ell^{3}$, where $\ell$ is the Planck length), each containing a 
spin which is normally rigidly enmeshed with the other spins, but which occasionally breaks loose.  Then,  
 for a duration equal to the Planck time $\tau$, it interacts both with a thermal bath,  
(characterized by the energy $\beta^{-1}$) and gravitationally 
with nearby particles (more on this shortly).  Following this brief dynamics it is 
frozen, no longer interacting with either bath or particles, once more enmeshed with other spins.
	
	Characterize the spin in cell i located at ${\bf x}_{i}$, which interacts over the interval $(t-\tau,t)$ and then freezes at time $t$, by $\sigma_{i,t}$ (a Pauli z-spin matrix) and mass $\mu\sigma_{i,t},$ where $\mu$ is the Planck mass.  Suppose that a spin-mass at ${\bf x}_{i}$ interacts gravitationally only with the particles in a sphere of radius $a$ around it,  and that the interaction is as if the particles are uniformly smeared over that sphere, with resulting mass density represented by the operator $\rho ({\bf x}_{i})$.  For simplicity,  take the particle Hamiltonian to vanish, so that $\rho$ is independent of $t$.  Then, since 
any initial statevector for the particles can be written as the sum of joint eigenstates of $\rho ({\bf x}_{i})$ 
for all ${\bf x}_{i}$,  calculations for the statevector need consider only the special case of one such term, for which $\rho ({\bf x}_{i})$ is a c-number, with the more general case being reconstructed afterward as the sum.  Then, for one such term,  the Hamiltonian for the gravitational interaction of all the spins which have 
interacted with the particles over the interval $(0,t)$ (up to a numerical factor which, for simplicity, is taken equal to 1)  is 
\begin{equation}\label{45}
H= -G\mu a^{2}\sum_{i,t'=\tau}^{t}\sigma_{i,t'}\rho ({\bf x}_{i}). 
\end{equation}

	Because all the spins act independently,  focus upon 
a group of spins excited over the time interval $(t', t'+\Delta t)$ which lie within a volume $\Delta V$.  Take $\Delta t$ and $\Delta V$ to be large ($\Delta t/\tau>>1$, $\Delta V/\ell^{3}>>1$), large enough to contain $N$ interacting spins where $N>>1$, but still small on the scale of particle physics (e.g., 
$\rho ({\bf x}_{i})$ is essentially constant over $\Delta V$).  Denote by $p$ the probability that 
a spin in a cell is freed up to interact over a time interval $\tau$ so, on average, 
\begin{equation}\label{46}
N=p\frac{\Delta V}{\ell^{3}}\frac{\Delta t}{\tau}.
\end{equation}
\noindent For simplicity, we shall take (46) to be the exact expression for $N$, not just for its average.   

	Now consider the (interaction picture) statevector $|\psi,t\rangle$.  It may be written as the direct product of  states associated with all volumes $\Delta V$ and all time intervals 
$\Delta t$  within $(0, t)$. The state which is the contribution of one such volume and interval has the form 
\begin{equation}\label{47}
|\chi\rangle=\sum_{s_{1}=-1}^{1}...\sum_{s_{N}=-1}^{1}|s_{1}...s_{N}\rangle|\hbox {bath},s_{1}...s_{N}\rangle C_{s_{1}...s_{N}}(\rho).
\end{equation}
\noindent The sum in Eq.(47) contains all the spin states entangled with the associated bath states which interacted with the spins and brought them to thermal equilibrium:  assume for simplicity that the bath states are mutually orthogonal.  
 
	Then, $|\chi\rangle$'s contribution to the density matrix describing the spins alone  is
\begin{equation}
\hbox{Tr}_{\hbox{bath}}|\chi\rangle\langle\chi |=\sum_{s_{1}=-1}^{1}...\sum_{s_{N}=-1}^{1}|s_{1}...s_{N}\rangle \langle s_{1}...s_{N}| |C_{s_{1}...s_{N}}(\rho)|^{2}.\nonumber 
\end{equation}
\noindent On the other hand, because these states are all in thermal equilibrium,  using Eq.(45)'s Hamiltonian for these spins, the thermal density matrix is:
\begin{equation}
\sum_{s_{1}=-1}^{1}...\sum_{s_{N}=-1}^{1}|s_{1}...s_{N}\rangle \langle s_{1}...s_{N}| 
\frac{e^{\beta G\rho \mu a^{2}\sum_{i=1}^{N}s_{i}}}{\sum_{s_{1}=-1}^{1}...\sum_{s_{N}=-1}^{1}e^{\beta G\rho \mu a^{2}\sum_{i=1}^{N}s_{i}}}.\nonumber 
\end{equation}	
Thus, by comparison of these two equations, 
\begin{equation}\label{48}
|C_{s_{1}...s_{N}}(\rho)|^{2}=
\frac{e^{\beta G\rho \mu a^{2}\sum_{i=1}^{N}s_{i}}}{\sum_{s_{1}=-1}^{1}...\sum_{s_{N}=-1}^{1}e^{\beta G\rho \mu a^{2}\sum_{i=1}^{N}s_{i}}}.
\end{equation}

	Eventually, $W$ associated with this space-time region will be taken $\sim S\equiv \sum_{i=1}^{N}\sigma_{i}$.  The statevector which is the sum of all states in (47) which have the same eigenvalue of $S$ 
and which is normalized to 1 is:  
\begin{equation}
|s\rangle\equiv\sum_{\{s_{i}\}, \sum s_{i}=s}|s_{1}...s_{N}\rangle|\hbox {bath},s_{1}...s_{N}\rangle \frac{C_{s_{1}...s_{N}}(\rho)}{\sqrt{\sum_{\{s_{i}\},\sum s_{i}=s}|C_{s_{1}...s_{N}}|^{2}}}\nonumber, 
\end{equation}
so, using (48), 
\begin{equation}\label{49}
\langle s|\chi\rangle=\bigg[\sum_{\{s_{i}\}, \sum s_{i}=s}|C_{s_{1}...s_{N}}(\rho)|^{2}\bigg]^{1/2}=
\bigg[\frac{\sum_{\{s_{i}\},\sum s_{i}=s}e^{\beta G\rho \mu a^{2}\sum_{i=1}^{N}s_{i}}}{\sum_{\{s_{i}\}}e^{\beta G\rho \mu a^{2}\sum_{i=1}^{N}s_{i}}}\bigg]^{1/2}.
\end{equation}	

	Evaluation of (49) is well known but, for completeness, it is sketched here.  The 
partition function in the denominator of (49) is (writing $C= G\rho \mu a^{2}$ and taking $N$ even)
\begin{equation}\label{50}
\hbox{Tr}e^{\beta CS} =\sum_{k=0}^{N}e^{\beta C(N-2k)}\frac{N!}k!{(N-k)!}=(2\cosh\beta C)^{N}, 
\end{equation}	
\noindent so the expression in (49) under the square root is 	
\begin{subequations}\label{51}
\begin{eqnarray}
\sum_{\{s_{i}\}, \sum s_{i}=s}|C_{s_{1}...s_{N}}(\rho)|^{2}&=&\frac{e^{\beta Cs}}{(2\cosh\beta C)^{N}}\frac{N!}{[(N-s)/2]![(N+s)/2]!}\\
&\approx&\frac{1}{\sqrt{2\pi N \cosh^{2}\beta C}}e^{-[s-N \tanh\beta C]^{2}/2N\cosh^{2}\beta C}\\
&\approx&\frac{1}{\sqrt{2\pi N}}e^{-[s-N\beta C]^{2}/2N}.
\end{eqnarray}
\end{subequations}		
\noindent In (51b) the well-known gaussian approximation to the binomial distribution has been employed\cite{Feller} 
which is valid for large $N$ and for s within a few standard deviations of its mean.  
The approximation in (51c) depends upon validity of the inequality
\begin{equation}\label{52}
\beta G\rho \mu a^{2}<<1
\end{equation}
\noindent which shall need to be checked.  

	Now it is possible to put together the contributions of all volumes $\Delta V$ and all time intervals 
$\Delta t$ within $(0, t)$ to construct the statevector $|\psi,t\rangle$.  
Define $|\tilde{s}\rangle\equiv\prod_{\Delta V, \Delta t}|s\rangle_{i,j}$,  the normalized basis vector which is the joint eigenvector of $S({\bf x}_{i},t_{j})$ for all ${\bf x}_{i}$ and for $0\leq t_{j}\leq t$.  The scalar 
product  $\langle \tilde{s} |\psi,t\rangle$ is the direct product of the expressions (49) so, from (51c) (without the normalization factors, which are tucked into the element of integration):
\begin{subequations}\label{53}
\begin{eqnarray}
\langle \tilde{s} |\psi,t\rangle&=&e^{-(4N)^{-1}\sum_{{\bf x}_{i},t_{j}=0}^{t}[s_{i,j}-N\beta G\rho \mu a^{2}]^{2}}\\
&=&e^{-4^{-1}N(\beta G \mu a^{2})^{2}\sum_{{\bf x}_{i},t_{j}=0}^{t}[s_{i,j}/N\beta G \mu a^{2}-\rho_{i}]^{2}}\\
&\approx&e^{-4^{-1}(p/\ell^{3}\tau)(\beta G \mu a^{2})^{2}\int d{\bf x}\int_{0}^{t}dt'[s'(\bf {x},t')-\rho(\bf {x})]^{2}},
\end{eqnarray}
\end{subequations}
\noindent where (53c) follows from (46), and we have defined:
\begin{equation}\label{54}
s'({\bf x},t)\equiv s_{i,j}/N\beta G \mu a^{2}.  
\end{equation}
	Now,  compare (53c) with CQC's comparable expression.  Note from (8) 
that $A({\bf x})=m_{0}^{-1}a^{3/2}\rho(\bf {x})$ (up to a numerical factor which shall be ignored) so the CQC expression comparable to Eq.(9) is
\begin{equation}\label{55}
\langle w |\psi,t\rangle=e^{-(4\lambda)^{-1}\int d{\bf x}\int_{0}^{t}dt'[w({\bf x},t')-2\lambda m_{0}^{-1}a^{3/2} \rho(\bf {x})]^{2}}.  
\end{equation}
\noindent  The model's Eq.(53c) and the CQC Eq.(55) are identical if   
\begin{subequations}\label{56}
\begin{eqnarray}
W({\bf x},t)&=&\frac{2\lambda a^{3/2}}{m_{0}}S'({\bf x},t)=\frac{2\lambda}
 {(\beta m_{0}c^{2})\ell a^{1/2}}\frac{S({\bf x},t)}{N},\\
4(\lambda \tau)(\ell/a)&=& p(\beta m_{0}c^{2})^{2}, 
 \end{eqnarray}
\end{subequations}	
\noindent where $G\mu=\ell c^{2}$ has ben used, and $S'{\bf x},t)$ is the  operator whose eigenvalues are $s'({\bf x},t)$ given in Eq.(54).

	Thus, this model produces CSL dynamics.  
	
	It remains to check whether the numerical values in Eqs.(52), (56) are reasonable.
Consider two possible temperatures for the bath, the cosmic radiation temperature $\beta^{-1}\approx 2\cdot 10^{-4}$ eV. and the Planck temperature $\beta^{-1}\approx 10^{28}$ eV.  For normal mass 
densities, $\rho\approx 1 \hbox{gm/cc}\approx 5\cdot 10^{33}\hbox{eV/cc}$, one gets
respectively $\beta G\rho \mu a^{2}\approx 10^{-6}$  and $10^{-38}$, so the inequality of 
Eq.(52) is satisfied for both.  

	For these two temperatures, (56b) gives $p\approx 10^{-112} $ and $10^{-49}$ respectively.  
The second number is too small for the conceptual picture we have outlined since, after 
$10^{49}\tau\approx 10^{5}$sec, the spin in every cell is likely to have interacted and frozen, so the 
process would cease.  However, one may change the conceptual picture.  It is possible for a spin in a cell to interact repeatedly, without changing the dynamical equation because the bath states 
 preserve the past orthogonality of the realizable states.   Although they are represented by bath states, they are labeled by the past values of W({\bf x},t).  
Each of these orthogonal realizable states at time $t$ evolves independently of the other states over the next $\Delta t$, via the CSL evolution.  

	Lastly, look at Eq.(56a).  Note that  $W\sim S/N$ is intensive, proportional to the mean 
value of the spin in a cell, so this result is independent of the size of $\Delta {\bf x}$, $\Delta t$ 
as it should be.  For the numerical coefficient in (56a), it is more informative to discuss $S'({\bf x},t)$'s relation to  $S({\bf x},t)$ rather than 
$W({\bf x},t)$'s relation to either,  because $S'({\bf x},t)$'s  mean value is simply  $\rho$ (see (53c)).  One obtains from Eq.(56a), for a cosmic bath and for a Planck bath respectively,  
$S'\approx (S/N)10^{39}$eV/cc and $S'\approx (S/N)10^{71}$eV/cc.  Both factors are large compared 
to normal mass density $\approx 10^{34}$eV/cc, so that the mean spin value $S/N$ 
 for such a density  is respectively $10^{-5}$ and $10^{-37}$.


\begin{thebibliography}{99}

\bibitem{Bohr} N. Bohr in  {\it Albert Einstein: Philosopher-Scientist}, 
edited by P. A. Schilpp (Harper and Row, New  York 1959), Chapter 7, p.201.

\bibitem{Peres-Fuchs} A. Peres and C. Fuchs, Physics Today {\bf 53}, 70 (2000).

\bibitem{Einsteinletter} A. Einstein in {\it Letters on Wave Mechanics}, 
edited by K. Przibram (Philosophical Library, New  York 1967), Letter 18, p.39.

\bibitem{Bell} J. S. Bell, in {\it Sixty-Two Years of Uncertainty},  edited by A. Miller (Plenum, New  York 1990), p. 17.

\bibitem{lev}For an overview and references, see L. Vaidman in {\it The Stanford Encyclopedia of Philosophy} edited by E. N. Zalta,  http://plato.stanford.edu/archives/sum2002/entries/qm-manyworlds/.
 
 \bibitem{zehzurek} E. Joos, H.D. Zeh, C. Kiefer D. Giulini, J. Kupsch, and I.-O. Stamatescu, {\it Decoherence and the Appearance of a Classical World in Quantum Theory}, (Springer, Heidelberg 2003);  W. H. Zurek, Rev. Mod. Phys {\bf 75}, 715 (2003)
 
\bibitem{griffithshartlegellmannomnes} R. E. Griffiths,  {\it Consistent Quantum Theory}, (Cambridge University Press, Cambridge 2003); R. Omnes,  {\it Understanding Quantum Mechanics}, Princeton University Press, Princeton 1999); M. Gell-Mann and J. B. Hartle in {\it Quantum  Classical  Correspondence: Proceedings of the  4th Drexel Symposium on Quantum Nonintegrability},  D. H.  Feng and B. L. Hu eds., (International Press 1997 ), p.3;  J. B. Hartle, {\it Linear Positivity and Virtual Probability}, quant-ph/0401108 and references therein. 

\bibitem{mekent} T. Breuer, Int. Journ. Theor. Phys. {\bf 32}, 2253 (1993); A. Kent, Phys. Rev. Lett.  {\bf 81}, 1982 (1998) and references therein;  P. Pearle in  {\it Quantum  Classical  Correspondence: Proceedings of the  4th Drexel Symposium on Quantum Nonintegrability},  D. H.  Feng and B. L. Hu eds., (International Press 1997 ), p.69; M. Schosshauer, Rev. Mod. Phys. {\bf 76}, 1267 (2004).  


\bibitem{me} P. Pearle, Phys. Rev. A {\bf 39}, 2277 (1989). 

\bibitem{GPR} G. C. Ghirardi, P. Pearle and A. Rimini, Phys. Rev. A {\bf 42}, 78 (1990). 

\bibitem{myreview} For a review of CSL, see P. Pearle in {\it  Open Systems and Measurement in
Relativistic Quantum Theory},  F. Petruccione and H. P. Breuer eds. (Springer Verlag, Heidelberg 1999), p.195.  

\bibitem{Ways}P. Pearle, Physical Review {\bf A48}, 913 (1993).

\bibitem{contmeas} A. Barchielli, L. Lanz and G. M. Prosperi, Nuovo Cimento {B 72}, 79 (1982) and  Found. Phys. {\bf 13}, 779 (1983);  R. L. Hudson and K. R. Parthasarathy, Comm. Math. Phys. {\bf 93}, 301 (1984);  C. M. Caves, G. J. Milburn, Phys. Rev. A. {\bf 36}, 5543 (1987); L. Diosi, Physics Letters A 132, 233 (1988); V. P. Belavkin, Physics Letters A 140, 355 (1989): L. Diosi, in Quantum Chaos-Quantum Measurement, edited by P. Cvitanovic et. al., (Kluwer, the Netherlands 1992), p. 299:   L. Accardi and I. V. Volovitch, quant-ph/9704029 and references therein on quantum white noise. 

\bibitem{Pearle'67} P. Pearle, Amer. Journ. Phys.  {\bf 35}, 742 (1967).

\bibitem{Schr} E. Schr\"odinger, "Die gegenw\"artige Situation in der Quantenmechanik", Naturwissenschaften 23: pp.807-812; 823-828; 844-849 (1935).  See the translation 
by J. D.Trimmer, in {\it  Quantum Theory and Measurement }, J.A. Wheeler and W.H. Zurek, eds., (Princeton University Press, New Jersey 1983).

\bibitem{GS} W. S. Gilbert and A. Sullivan, "And we are right, I think you'll say, To argue in this kind of way; And I am right, And you are right, And all is right---too-looral-lay!"  in {\it The Mikado},  (1885).

\bibitem{kent} A. Kent, Phys. Scripta {\bf T76}, 78 (1998).

\bibitem{exptlpapers} B. Collett, P. Pearle, F. Avignone and S. Nussinov, 
 Found. Phys. {\bf 25}, 1399 (1995); P. Pearle, J. Ring, J. I. Collar and F. Avignone,
Found. Phys. {\bf 29}, 465 (1999); G. Jones, P. Pearle and J. Ring, Found. Phys. {\bf 34}, 1467 (2004).

\bibitem{energy}P. Pearle, Found. Phys. {\bf 30}, 1145 (2000).

\bibitem{Adler} S. L. Adler, {\it Quantum Theory as an Emergent Phenomenon} (Cambridge University Press, Cambridge UK 2004).  

\bibitem{decoh} E. Joos and H. D. Zeh, Z. Phys. B: Condensed Matter  {\bf 59}, 223 (1985).  

\bibitem{gravity} F. Karolyhazy, Nuovo Cimento {\bf42A}, 1506 (1966); 
F. Karolyhazy, A Frenkel and B. Lukacs 
in {\it Quantum Concepts in Space and Time}, edited by R. Penrose and 
C. J. Isham (Clarendon, Oxford 1986), p. 109; 
R. Penrose, Ibid, p. 129; in {\it The Emperor's New Mind}, (Oxford University Press, Oxford, 1992),  in {\it 
Shadows of the Mind}, (Oxford University Press, Oxford, 1994) and in Gen. Rel. and Grav.,  {\bf 28}, 581 (1996);  L. Diosi, {\it Phys. Rev. A} {\bf 40}, 1165 (1989);  
G. C. Ghirardi, R. Grassi and A. Rimini, {\it Phys. Rev. A} {\bf 42}, 1057 (1990);  

\bibitem{pearlesquires}P. Pearle and E. Squires, {\it Found. Phys.} {\bf 26}, 291 (1996).

\bibitem{smolinseth}S. Major, Am. J. Phys. {\bf 67}, 972 (1999) and references therein.

\bibitem{GRW} G. C. Ghirardi, A. Rimini and T. Weber, Phys. Rev. D {\bf 34}, 470 (1986); 
Phys. Rev. D {\bf 36}, 3287 (1987); Found. Phys. {\bf 18}, 1, (1988). 

\bibitem{tachyon} J. E. Murphy, Phys. Rev. D{\bf 6}, 426, (1972). 

\bibitem{super}D. Giulini in {\it  Relativistic Quantum Measurement and 
Decoherence},  F. Petruccione and H. P. Breuer eds. (Springer Verlag, Heidelberg 2000), p.67, and references therein.  

\bibitem{nonwhite} P. Pearle, Phys. Rev. A {\bf 48}, 913 (1993), in {\it Stochastic Evolution of Quantum States in Open Systems and in Measurement Processes}, edited by L. Diosi 
and B. Lukacs (World Scientific, Singapore 1994), p. 79, 
in {\it Perspectives on Quantum Reality}, edited by R. Clifton (Kluwer, Dordrecht 1996), p. 93, in  Phys. Rev. A{\bf 59}, 80 (1999); A. Bassi and G.C. Ghirardi, Phys. Rev. A{\bf 65}, 042114 (2002).

\bibitem{timeref} For an example of a time operator conjugate to a semi-bounded Hamiltonian, see 
J. G. Muga, C. R.Leavens and  J. P. Palao, Phys. Rev. A{\bf 58}, 4336, (1998) (and references therein). This  paper discusses Aharanov and Bohm's arrival time operator $T_{AB}\equiv-(m/2)[xp^{-1}+ p^{-1}x]$, conjugate to the free particle  Hamiltonian $p^{2}/2m$.  Typical of such behavior,  $T_{AB}$ is symmetric ($T_{AB}^{\dagger} =T_{AB}$ on the domain of $T_{AB}$), but not self-adjoint (the domain of $T_{AB}^{\dagger}$ is larger than the domain of $T_{AB}$) and, as a consequence, the eigenvectors of 
$T_{AB}$ for different eigenvalues are not othogonal.

\bibitem{exp}P. Pearle, Phys. Rev. D{\bf 29}, 235 (1984); ref. \cite{exptlpapers}; P. Pearle and B. Collett, Found. of Phys. {\bf 33}, 1495 (2003); ref. \cite{Adler}; W. Marshall, C. Simon, R. Penrose and D. Bouwmeester, Phys. Rev. Lett.{\bf 91}, 130401 (2003);  A. Bassi, E. Ippoliti and S.L. Adler, Phys. Rev. Lett. {\bf 94}, 030401 (2005)


\bibitem{Feller} W. Feller, {\it An Introduction to Probability Theory and its Applications, Vol. 1} (Wiley, New York 1950).


\end{thebibliography}
\end{document}